\documentclass[apj]{emulateapj}
\shorttitle{MHD Wind Dynamics}
\shortauthors{Cui $\&$ Yuan 2019}
\usepackage{amsmath}
\usepackage{hyperref}
\usepackage{physics}
\hypersetup{colorlinks=true, citecolor = blue, urlcolor = blue, linkcolor=blue}

\begin{document}
\title{Large-scale dynamics of winds originated from black hole accretion flows: (II) Magnetohydrodynamics}
\author{Can Cui$^{1,2}$ and Feng Yuan$^{1,2}$}
\affil{$^{1}$Shanghai Astronomical Observatory, Chinese Academy of Sciences, Shanghai 200030, China\\
$^{2}$University of Chinese Academy of Sciences, 19A Yuquan Road, Beijing 100049, China\\
\href{mailto:ccui@shao.ac.cn}{ccui@shao.ac.cn}; \href{mailto:fyuan@shao.ac.cn}{fyuan@shao.ac.cn}}

\begin{abstract}

The great difference in dynamical range between small-scale accretion disk simulations and large-scale or cosmological simulations places difficulties in tracking disk wind kinematics. In the first paper of this series, we have studied dynamics of hydrodynamic winds from the outer edge of the accretion disk towards galactic scales. In this paper, we further incorporate magnetic fields by employing one-dimensional magnetohydrodynamic (MHD) model, with fiducial boundary conditions set for hot accretion flows. 
The wind solution is achieved through requesting gas to pass through the slow, Alfv\'{e}n and fast magneto-sonic points smoothly. Beyond the fast magneto-sonic point, physical quantities are found to show power-law dependences with cylindrical radius $R$, i.e. $\rho \propto R^{-2}, v_{\rm p}\propto {\rm const.}, v_{\rm \phi}\propto R^{-1}, B_{\rm \phi}\propto R^{-1},$ and $ \beta \propto \rho^{\gamma-1}$. 
The magnetization of wind is dominant in determining the wind properties. The wind is accelerated to greater terminal velocities with stronger magnetizations. The fiducial parameters result in a terminal velocity about $0.016c$.
The dependance of wind physical quantities on temperature, field line angular velocity, and adiabatic index is also discussed. 

\end{abstract}
\section{Introduction}

It is widely accepted that disk winds are broadly present in black hole accretion systems. On small scales, they serve as an indispensable ingredient of black hole accretion, determining the density and temperature of accretion flows, subsequently affecting the emitted spectrum (e.g., \citealp{yuan03}). On large scales, they are key to interactions and coevolution of the central black hole and its host galaxy (e.g., \citealp{ciotti_etal10,ciotti_etal17,ostriker_etal10,choi12,weinberger_etal17,eisenreich_etal17,yuan2018,yoon2018,yoon2019}). Wind launching mechanisms are extensively studied in the literature. In particular, three mechanisms have been proposed, namely, the thermally driven (e.g., \citealp{Begelman1983,font_etal04,luketic_etal10,wp12}), the radiation driven (e.g. \citealp{Murray1995,Proga2000,Proga2004,Nomura2017}), and the magnetically driven (e.g., \citealp{bp82,lynden96,lynden03}). 

The magnetohydrodynamic (MHD) wind theory has been long established, inherited by the seminal work of \citet{bp82} and \citet{lynden96,lynden03}, following by intense studies over the last few decades (e.g. \citealp{pn83,pn86,sakurai1985,sakurai87,konigl89,lbc91,pp92,cl94,cs94,fp95,li95,li96,ferreira97,ostriker97,vlahakis00,everett05,fukumura_etal10,bai_etal16}). Magnetically driven winds can be generally categorized into two classes. One of which is the magneto-centrifugal winds where the poloidal magnetic field dominates, and the other is magnetic pressure gradient driven winds where the toroidal field dominates. 

A centrifugal force is able to drive winds if the poloidal component of magnetic fields makes an angle more than $30^\circ$ from the rotational axis \citep{bp82}. The launching of magneto-centrifugal winds generally requires the presence of a large-scale, ordered magnetic field threading the disk with a poloidal component at least comparable to the toroidal magnetic field (e.g., \citealp{cp88,pp92}). Global MHD simulations with time-dependency have been performed to study the structure and evolution of these winds, though the internal structure of the disk is usually ignored with winds being ejected at the boundary (e.g., \citealp{ust_etal95,ust_etal99,romanova_etal97,op97a,op97b,op99,klb99,kks02,anderson_etal05,pro06,zanni_etal07,pf10}). The toroidal magnetic field builds up due to disk rotation, giving rise to winds driven by magnetic pressure gradient \citep{lynden96,lynden03}. Depending on the ratio of poloidal to toroidal field strength, the wind will transition from magneto-centrifugally driven to magnetic pressure gradient driven along its propagation \citep{us85,pn86,su86,sn94,contopoulos95,ks97,op97b}.

Global simulations on cold accretion disks suffer from proper implementation of radiative transfer processes, which is key to the thin disk model. Moreover, the simultaneously modelling of geometrically thin disks with resolved  gas dynamics and propagation of disk winds to large radii would be prohibitively time-consuming. Previous numerical studies generally do not resolve the full internal structure of the disk. Instead, they employ simplifications by injecting winds from the simulation boundary, assuming specific wind driving mechanism(s) (e.g., \citealp{Proga2000,pk02,luketic_etal10}). Since the simplified model is not able to generate wind self-consistently from accretion disks, the wind properties obtained are not fully reliable. On the other hand, the theoretical understanding of winds launched from hot accretion flows is more advanced, partly due to the radiation is dynamically unimportant in hot accretion flows and to the ease of simulating geometrically thick flows. The early speculation of strong winds existing in hot accretion flows \citep{narayan1994,blandford1999} was later confirmed by numerical simulations \citep{yuan_etal12a,yuan_etal12b,narayan_etal12,li_etal13}.   

Winds from hot accretion flows have been thoroughly studied in \citet{yuan_etal15} (hereafter Y15). They analyze data from 3D general relativistic (GR) MHD simulations via a virtual particle trajectory approach, which effectively discriminates real wind from turbulent flows. Winds originating from smaller radii are found to have larger poloidal velocities, and the velocity roughly keeps constant during the outward propagation. Differentiating from global simulations of thins disks, winds are self-consistently generated in hot accretion flow simulations with the internal dynamics of accretion flows resolved so that reliable wind properties are obtained. 

The simulations mentioned above can only track winds on accretion disk scales. Nevertheless, wind properties beyond this scale are of great importance in order to understand its role in the interactions between active galactic nuclei (AGN) and host galaxies. Recent cosmological simulations invoke winds from hot accretion flows interacting with the interstellar medium on galactic scales to overcome serious problems in galaxy formation, e.g., reducing star formation efficiency in the most massive halos (e.g., \citealp{weinberger_etal17}). Moreover, \citet{yuan2018} comprehensively include feedback by wind and radiation from AGNs in cold and hot feedback modes and find that wind plays a dominant role in both modes, though radiative feedback cannot be neglected.

The dynamics of disk winds have been studied in the context of black hole accretion disks with most devotions on thin disks (e.g., \citealp{cl94,romanova_etal97,Proga2000,proga03,Proga2004,luketic_etal10,wp12,Cao2014,
ca16,Nomura2017,waters18}), and some of these works have extended to large radii. In this series of work, we aim to study the wind dynamics beyond accretion disk scales via analytical method. A hydrodynamic model has been adopted in our first paper to study thermally driven winds \citep{cyl19}. In this paper, we employ one-dimensional MHD equations to understand how magnetic fields influence the wind dynamics with special attention to those from hot accretion flows. The key factor of studying the large-scale wind dynamics lies in the precise adoption of boundary conditions because the MHD equations controlling the wind dynamics are a set of differential equations. In this work, we will revisit the large-scale dynamics with realistic boundary conditions from small-scale accretion disk simulations and focus on winds from hot accretion flows.

Analytical studies of magnetized winds in cold black hole accretion disks have been conducted in the literature. Some of these works invoke the simplification of self-similarity in solving MHD equations hence suffer from the fact that boundary conditions are not needed to be prescribed (e.g., \citealp{everett05,fukumura_etal10}). Despite the rarity of large-scale wind studies from hot accretion flows, a recent work by \citet{bm18} investigates the wind properties of Advection-dominated accretion flows via resistive MHD equations. However, their results are also limited by the adoption of self-similar solutions. In this work, we pursue study on magnetized disk winds by the standard Weber \& Davis model and solve the set of MHD equations self-consistently, with the the most realistic boundary conditions taken from small-scale accretion disk simulations (Y15). 

The paper is organized as follows. We describe the MHD wind model and the analytical approach in \S\ref{sec:model}. In \S\ref{sec:bc}, we discuss the boundary conditions in terms of hot accretion flows and thin disks. We present solutions by detailing the magnetization, temperature, mass loading, and acceleration mechanism of wind in \S\ref{sec:result}. Parameter studies on adiabatic indices and disk angular velocities are conducted in \S\ref{sec:parastudy}. Finally, we summarize the main findings and discuss the results in \S\ref{sec:cd}.

\section{Model Description and Equations}\label{sec:model}

The steady ($\partial/\partial t=0$), axisymmetirc ($\partial/\partial \phi=0$) model of magnetized disk winds is presented in this section, following equations introduced in \citet{wd67}. The wind geometry is prescribed in \S\ref{sec:geo}. The set of equations to be solved is described in \S\ref{sec:eqn} with critical points properties detailed in \S\ref{sec:critical}. The numerical procedures in solving MHD equations are elaborated in \S\ref{sec:procedure}, and a sample solution is displayed in \S\ref{sec:solutionplane}. We list the physical quantities in \S\ref{sec:pred} which will facilitate the analysis. 

\subsection{Wind Geometry}\label{sec:geo}
Using cylindrical coordinates ($R,\phi,z$), we decompose the magnetic field $\mathbf{B}$ and velocity field $\mathbf{v}$ at any point in the outflow into poloidal and toroidal components,
\begin{equation}
\textbf{B} = \textbf{B}_\textrm{p} + B_{\phi} \hat{\phi}, \quad   \textbf{v} = \mathbf{v}_\textrm{p} + v_{\phi} \hat{\phi}, 
\end{equation}
where $\mathbf{v}_\mathrm{p}$, $\mathbf{B}_\mathrm{p}$ are the poloidal velocity and magnetic field components, $v_\phi$ is the rotational velocity, and $B_\phi$ is the toroidal magnetic field component. 

The wind is prescribed to be launched from disk surface at ($R_0,z_0)$, known as the wind base or the footpoint of magnetic field line. Assuming large poloidal filed lines threading the accretion disk, the field line is anchored at wind base and is taken to be straight for $R>R_0$ in the poloidal plane. This simplified assumption enables us to easily incorporate with the wind geometry, and it is valid through small-scale accretion disk simulations for hot accretion flows (Y15, see their Figure 1). With a constant inclination angle $\theta$ to the rotational axis, we parametrize the poloidal filed line, which is also the streamline of the wind due to flux freezing, by $R=R_0+s\cos\theta$ and $z=z_0+s\sin\theta$, where $s$ denotes the length along the poloidal magnetic field.

The 1D Weber and Davis model requires the prescription of poloidal field strength along the streamline. We adopt the divergence free condition, writing the function of $B_\mathrm{p}(R)$ in the form  
\begin{equation}
B_\mathrm{p}(R) = B_\mathrm{p0}\left(\frac{R}{R_0}\right)^{-2}, 
\label{eq:bp}
\end{equation}
where subscript naught denotes quantities at the wind base.

\subsection{Conservation Laws}\label{sec:eqn}
A magnetized disk wind is described by six equations for six variables, the gas density $\rho$, pressure $P$, poloidal components of velocity and magnetic field $v_\mathrm{p}$, $B_\mathrm{p}$, and toroidal components $v_\phi$, $B_\phi$. One of these equations prescribing the strength of poloidal magnetic field along the streamline is shown in Equation \eqref{eq:bp}. Another among these is the polytropic equation of state
\begin{equation}
P=K\rho^\gamma, \label{eq:eos}   
\end{equation}
where $K$ and $\gamma$ are constants with latter representing the polytropic index. The sound speed is defined by $c_{\rm s}^2 \equiv \partial P/\partial \rho= \gamma P/\rho$. The polytropic relation is employed to express the enthalpy term in conservation of specific energy (Equations \ref{eq:e} and \ref{eq:h}).  

The rest four (Equations \ref{eq:kappa}-\ref{eq:e}) are conservation laws derived from stationary ideal MHD. In the Gaussian unit system, these equations read \citep{spruit96}
\begin{gather} 
\nabla \cdot (\rho \textbf{v})=0,  \label{eq:cont}   \\ 
\rho(\textbf{v} \cdot \nabla )\textbf{v}= - \nabla P - \rho \nabla \Phi + \frac{1}{4\pi}(\nabla \times \textbf{B})\times \textbf{B}, \label{eq:mt}\\
\nabla \times (\textbf{v} \times \textbf{B})  = 0, \label{eq:induction}  \\ 
\nabla \cdot \textbf{B} = 0, \label{eq:solenoidal}
\end{gather}
where $\Phi$ represents the gravitational potential. Equation \eqref{eq:cont} is the continuity equation, and Equation \eqref{eq:mt} is the equation of motion. Equations \eqref{eq:induction} and \eqref{eq:solenoidal} are the induction equation and the divergence free condition which states that no magnetic monopoles exists. Due to axisymmetry and the conservation of magnetic flux, the poloidal magnetic field is derived from the magnetic flux function $\psi$ by \citep{spruit96,ogilvie16}
\begin{equation}
\mathbf{B}_\mathrm{p}=\frac{1}{R}\nabla \psi \times \hat{\phi}. \label{eq:flux}   
\end{equation}
Hence we have $\mathbf{B}\cdot \nabla \psi=0$, which indicates that the flux function $\psi$ labels field lines or their surfaces of evolution. For steady and axisymmetric flow, Equations (\ref{eq:cont})-(\ref{eq:solenoidal}) are reduced to four conservation laws with four invariants $\kappa, \omega, l, \varepsilon$. These quantities are functions of $\psi$, thereby conserved along each individual field line. 

The first of these invariants can be derived from the continuity equation:
\begin{equation}
\kappa(\psi) \equiv \frac{4\pi\rho v_\mathrm{p}}{B_\mathrm{p}},  \label{eq:kappa}   
\end{equation}
where $\kappa$ is the ratio of mass flux to magnetic flux. The induction equation further gives the conservation of angular velocity of the filed line:
\begin{equation}
\omega(\psi) \equiv \frac{v_{\phi}}{R} - \frac{\kappa B_\phi}{4\pi\rho R}, \label{eq:omega} 
\end{equation}
where $v_\phi=\Omega R$, and $\Omega$ is the gas angular velocity. Following Equation \eqref{eq:induction}, the steady and axisymmetric conditions give $\mathbf{v}_\mathrm{p}\times \mathbf{B}_\mathrm{p}=0$, since they do not allow the existence of toroidal electric field ($\mathbf{E_\phi=-v_\mathrm{p}\times B_\mathrm{p}}/c=0$). Thus, the poloidal velocity and magnetic field are everywhere parallel to each other, $\mathbf{v}_\mathrm{p} \parallel  \mathbf{B}_\mathrm{p}$. This expresses the flux freezing condition and does not depend on the reference frames. Along with Equation \eqref{eq:omega}, the total gas velocity is parallel to the total magnetic field in the frame rotating with $\omega$. The azimuthal component of the equation of motion implies the conservation of angular momentum on each filed line:
\begin{equation}
l(\psi) \equiv  v_{\phi} R - \frac{RB_\phi}{\kappa}. \label{eq:l} 
\end{equation}
The first term on the right is the ordinary specific angular momentum, and the second term represents the torque associated  with the magnetic stresses. The Bernoulli integral or the conservation of specific energy expresses the last invariant:
\begin{equation}
\varepsilon(\psi) \equiv  \frac{1}{2}v^2_\mathrm{p}+\frac{1}{2}(v_\mathrm{\phi}-\omega R)^2+\Phi_\mathrm{eff}+h,
\label{eq:e}
\end{equation}
where $\Phi_\mathrm{eff}=\Phi-\omega^2R^2/2$ is the combined potential energy of centrifugal and gravitational forces along the field line. The gravitational potential by a pointmass is defined as $\Phi=-GM_\mathrm{BH}(R^2+z^2)^{-1/2}$, where $M_\mathrm{BH}$ is the mass of the central black hole. Along with the polytropic law, the enthalpy $h=\int dP/\rho$ is written as  \begin{equation}
h=\frac{\gamma}{\gamma-1}K\rho^{\gamma-1}.
\label{eq:h}
\end{equation}
It is obvious from the angular velocity and centrifugal potential terms that Equation \eqref{eq:e} is written in the rotating frame with footpoint angular velocity $\omega$. In the rest frame, the Bernoulli constant is given by $\tilde{\varepsilon}=v^2/2+\Phi+h-R\omega B_\phi/\kappa$, where $v=(v^2_\mathrm{p}+v^2_\phi)^{1/2}$. The last term corresponds to the Poynting flux and is not shown in $\varepsilon$. This is because in the rotating frame, the magnetic field is strictly parallel to the flow velocity such that the Lorentz force $\mathrm{F_L}=\mathbf{J}\times\mathbf{B}/c$ everywhere perpendicular to $\mathbf{B}$ is perpendicular to $\mathbf{v}-\omega R\hat{\phi}$, hence the field does no work in this frame. The two Bernoulli integrals are related by $\varepsilon=\tilde{\varepsilon}-l\omega$.

\subsection{Critical Points}\label{sec:critical}

It is convenient to introduce the poloidal Alfv\'{e}nic Mach number $M_\mathrm{A}$, which is defined as
\begin{equation}
M^2_\mathrm{A} = \frac{v^2_\mathrm{p}}{v^2_\mathrm{Ap}} = \frac{\kappa^2}{4\pi\rho},
\label{eq:MA}
\end{equation}
where $v_\mathrm{Ap}= B_\mathrm{p}/\sqrt{4\pi \rho}$ is the poloidal Alfv\'en velocity. Eliminating $B_\phi$ in Equations \eqref{eq:omega} and \eqref{eq:l} gives 
\begin{equation}
v_\phi=\frac{M^2_\mathrm{A}l/R-\omega R}{M^2_\mathrm{A}-1}.
\label{eq:vphi}
\end{equation}
The radius $R=R_\mathrm{A}$ where $M_\mathrm{A}=1$ is the Alfv\'{e}nic point. The denominator of the expression for $v_\phi$ goes to zero at this point, hence we require the numerator to vanish identically. This results in a simple expression for the conserved specific angular momentum as 
\begin{equation}
l=\omega R^2_\mathrm{A}.
\label{eq:l2}
\end{equation}

Substituting Equations (\ref{eq:kappa})-(\ref{eq:l}) into Equation (\ref{eq:e}) the Bernoulli integral, we can express $\varepsilon$ as a function of $\rho$ and $R$
\begin{equation}
\varepsilon = H(\rho,R),
\end{equation}
and $H(\rho,R)$ takes the explicit expression
\begin{equation}
H(\rho, R)= \frac{1}{2}\frac{\kappa^2B_{\rm p}^2}{(4\pi\rho)^2} + \frac{1}{2}\frac{(l/R-\omega R)^2\kappa^4}{(4\pi\rho-\kappa^2)^2}+\Phi_\mathrm{eff}+h. 
\end{equation}
Substituting $\rho$ by $M_\mathrm{A}$ through Equation \eqref{eq:MA}, we obtain
\begin{equation}
H(M_\mathrm{A}, R) = \frac{1}{2}\frac{M^4_\mathrm{A}B_{\rm p}^2}{\kappa^2} + \frac{1}{2}\frac{(l/R-\omega R)^2}{(M^{-2}_\mathrm{A}-1)^2}+\Phi_\mathrm{eff}+h. 
\label{eq:MA}
\end{equation}
Critical points can be obtained via requiring the partial derivatives of $H(\rho,R)$ to be zero,
\begin{equation}
\frac{\partial H(\rho,R)}{\partial \rho}=\frac{\partial H(\rho,R)}{\partial R}=0,
\label{eq:partial}
\end{equation}
which gives the slow ($\rho_{\rm s}, R_{\rm s}$) and the fast ($\rho_{\rm f}, R_{\rm f}$) magneto-sonic point. The subscripts $s$ and $f$ denote slow and fast critical points. In particular, these points manifest themselves in the partial derivative of $\rho$,
\begin{align}
\rho \frac{\partial H}{\partial \rho}  &= -v_{\rm p}^2 -\frac{(\Omega-\omega)^2R^2}{1-M^2_\mathrm{A}} + c_\mathrm{s}^2  \nonumber \\
 &= \frac{v_\mathrm{p}^4-(c_\mathrm{s}^2+v_\mathrm{Ap}^2+v_\mathrm{A\phi}^2)v_\mathrm{p}^2+c_\mathrm{s}^2v_\mathrm{Ap}^2}{v_\mathrm{Ap}^2 - v_\mathrm{p}^2}  \nonumber \\
 &= \frac{(v_\mathrm{p}^2-v_\mathrm{sp}^2)(v_\mathrm{p}^2-v_\mathrm{fp}^2)}{v_\mathrm{Ap}^2 - v_\mathrm{p}^2},   \label{eq:prho} 
\end{align}
where $v_\mathrm{A\phi}= B_\phi/\sqrt{4\pi \rho}$ is the toroidal Alfv\'en velocitiy, and the square of the sound speed is defined as $c^2_\mathrm{s}\equiv \partial P/\partial \rho=\gamma P/\rho$. The expressions of $v_\mathrm{sp}$ and $v_\mathrm{fp}$ are given by the quadratic formula 
\begin{equation}
v_\mathrm{sp}^2,v_\mathrm{fp}^2 = 
 \frac{(c_\mathrm{s}^2+v_\mathrm{A}^2)\mp \sqrt{(c_\mathrm{s}^2+v_\mathrm{A}^2)^2 - 4c_\mathrm{s}^2v_\mathrm{Ap}^2}}{2},
\end{equation}
where $v_\mathrm{A}^2=v_\mathrm{Ap}^2+v_\mathrm{A\phi}^2$. 
The left hand side vanishes in Equation \eqref{eq:prho} when $v_\mathrm{p}$ equals either the slow mode velocity $v_\mathrm{sp}$ or the fast mode velocity $v_\mathrm{fp}$. 
Besides the conditions for critical points imposed by Equation \eqref{eq:partial}, a further constraint is placed to equate the energies at slow and fast points to the Bernoulli constant  
\begin{equation}
H(\rho_{\rm s}, R_{\rm s}) =\varepsilon, \qquad H(\rho_{\rm f}, R_{\rm f}) =\varepsilon.
\label{eq:critE}
\end{equation}

We have seen that at the slow and fast magneto-sonic points six equations are introduced, i.e. Equations \eqref{eq:partial} and \eqref{eq:critE} at $(\rho_{\rm s}, R_{\rm s})$ and $(\rho_{\rm f}, R_{\rm f})$, while eight variables ($\rho_{\rm s}, R_{\rm s}, \rho_{\rm f}, R_{\rm f}, \kappa, \omega, l, \varepsilon$) are presented. Among these variables, if two are specified then the remaining six can be determined. In this paper, we fix $\omega$ and $\varepsilon$, and solve for the rest variables $\rho_{\rm s}, R_{\rm s}, \rho_{\rm f}, R_{\rm f}, \kappa$ and $l$. Hence, any wind solution is characterized by $\omega$ and $\varepsilon$. 

\subsection{Numerical Procedures}\label{sec:procedure}

We have verified that in pure hydrodynamic model wind solutions should be either supersonic or transonic, whereas the subsonic solutions are not likely to exist due to low-frequency acoustic perturbations \citep{cyl19}. Thereby, our MHD equations are solved under the condition that all solutions should pass through the slow, Alfv\'{e}n, and fast magneto-sonic point smoothly. The general equations to be solved are given in Equations \eqref{eq:partial} and \eqref{eq:critE}. These equations shall be fulfilled at the slow and fast magneto-sonic point. To start with, we iterate over a set of values for $l$ and solve for $\kappa, R_{\rm s}, M_{\rm As}^2$ at the slow critical point. We then solve for  $l, R_{\rm f}, M_{\rm Af}^2$ at the fast critical point using $\kappa$ found at the slow point. The initial guess of $l$ and its computed value at fast point are compared for each iteration until the two values match. The initial guesses for all these variables are taken through the inspection of the contour $H(\rho, R)=\varepsilon$. Once $\rho_{\rm s}, R_{\rm s}, \rho_{\rm f}, R_{\rm f}, \kappa$ and $l$ are found, the rest variables can be obtained via either solving $H(\rho, R)=\varepsilon$ directly or tracking along the contour curve. 

Success in finding the solution at the first time involves difficulties. However, once the first set of solution is achieved, it can be used as the initial guess as one alters the parameters to find new sets of solutions. Note that some solutions have their initial poloidal velocities exceed the slow mode velocities, or equivalently $R_{\rm s}<R_0$. The wind can be accelerated through any other mechanisms when $R<R_0$ which is outside the scope of this paper, and we focus on the large-scale wind dynamics for $R>R_0$ in the ideal MHD regime as the model established in this paper.  

\subsection{Solution plane}\label{sec:solutionplane}

\begin{figure}[ht!]
\epsscale{1.1}
\plotone{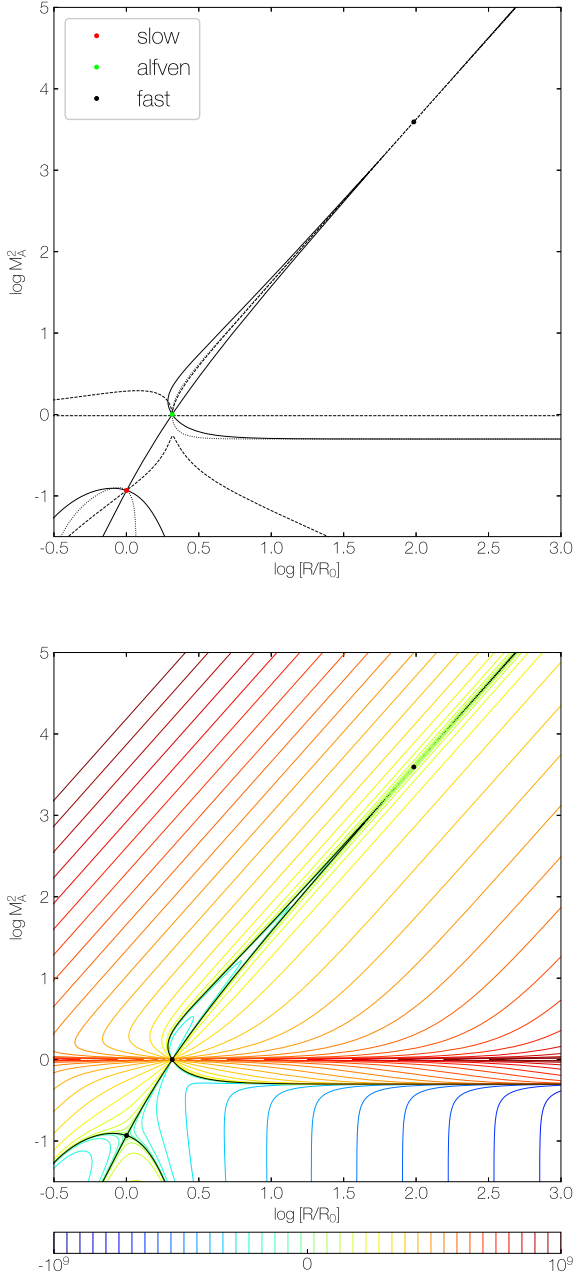}
\caption{Solution plane of the wind model. Top panel: The solid, dashed, and dotted curves represent the solutions of $H(\rho, R)=\varepsilon$, $\partial H(\rho,R)/\partial \rho=0$, and $\partial H(\rho,R)/\partial R=0$. The intersections between these curves show the loci of the slow (red), Alfv\'{e}n (green), and fast (black) mode velocities, respectively, from the bottom to the top of the $(M_\mathrm{A}^2, R)$-plane. Bottom panel: solution curves by drawing contours of $H(\rho, R)$ with Equation \eqref{eq:MA}. Colors delineate deviations from $H(\rho, R)=\varepsilon$, for which is highlighted in black. The wind solution should be part of the solid black curve that smoothly connects the slow, Alfv\'{e}n, and fast points.}
\label{fig:contours}
\end{figure}

Figure \ref{fig:contours} shows the solution plane of the wind model. In the top panel, we plot curves of  $H(\rho, R)=\varepsilon$, $\partial H(\rho,R)/\partial \rho=0$, and $\partial H(\rho,R)/\partial R=0$ in the $(M_\mathrm{A}^2, R)$-plane. Their interactions represent the slow, Alfv\'{e}n, and fast magneto-sonic points, respectively. The wind solution curve is part of the solid curve that smoothly connects the slow, Alfv\'en and fast magneto-sonic point in order, with poloidal flow velocity exceeds the slow,  Alfv\'{e}n, and fast mode velocities when the solution crosses the corresponding critical points.

In the bottom panel, we plot the contours of $H(\rho, R)$ and highlight the $H(\rho, R)=\varepsilon$ curve in black, where our wind solution resides in, using Equation \eqref{eq:MA}. The colors delineate contours with Bernoulli integrals that deviate from $\varepsilon$. Inspecting on the color contours near the critical points, it shows that the slow and fast critical points are saddle points. Although the Alfv\'{e}n point is a focus of a bundle of cruves, it does not impose additional constraints on the wind solution. This can be understood as the condition has already been applied at the Alfv\'{e}n point in deriving Equation \eqref{eq:l2}. Strictly speaking, it should be referred to Alfv\'{e}n point but not Alfv\'{e}n critical point.

\subsection{Definition of Physical Quantities}\label{sec:pred}

The Alfv\'{e}n point separates the wind solution into two regimes in terms of rotation. For $M_\mathrm{A} \ll 1$ Equation \eqref{eq:vphi} gives 
\begin{equation}
v_\phi \approx \omega R,
\label{eq:fieldgas1}
\end{equation}
where the fluid corotates ($\Omega=\omega$) with the angular velocity of the magnetic field lines. For $M_\mathrm{A} \gg 1$,
\begin{equation}
v_\phi \approx \frac{l}{R},
\label{eq:fieldgas2}
\end{equation}
and the fluid rotates by conserving its specific angular momentum. In practice, wind usually starts from low velocities ($M_\mathrm{A} \ll 1$) so that we can identify $\omega \approx \Omega_0$, where $\Omega_0$ is the angular velocity of wind at the footpoint. The specific angular momentum is $\Omega_0R_0^2$ at the launching point. Once the wind is accelerated to the Alfv\'en point, the excess of the specific angular momentum is $\Omega_0(R_{\rm A}^2-R_0^2)$. By the assumption that $\Omega_0 \approx \Omega_{\rm K}(R_0)$, the wind mass loss rate is related to the mass accretion rate inside the disk by \citep{fp95,bai_etal16}
 \begin{equation}
\xi = \frac{1}{\dot{M}_\mathrm{acc}} \dv{\dot{M}_\mathrm{wind}}{ \ln R} = \frac{1}{2} \frac{1}{(R_\mathrm{A}/R_0)^2-1},
\label{eq:xi}
\end{equation}
where $\dot{M}_\mathrm{acc}$ is the wind-driven accretion rate, $\dot{M}_\mathrm{wind}(R)$ is the cumulative mass-loss rate, $\xi$ is called the ejection index, and the ratio $R_\mathrm{A}/R_0$ is often referred to the magnetic lever arm. The location of the Alfv\'{e}n point can thereby provide a convenient measure of the mass loss to the accretion rate. 

To quantify the wind properties, we introduce the dimensionless mass loading parameter $\mu$, defined as
\begin{equation}
\mu = \kappa \frac{\omega R_0}{B_{\rm p0}}.
\label{eq:mu}
\end{equation} 
The invariant $\kappa$ in Equation \eqref{eq:kappa} represents the mass flux per field line. The mass loading parameter is obtained by normalizing $\kappa$ with $B_{\rm p0}/ \omega R_0$. The wind is lightly loaded when $\mu \ll 1$, and heavily loaded when $\mu \gg 1$.
Characteristic quantities of wind can be written explicitly as a function of $\mu$ in a simplified model (Equations \ref{eq:RA}-\ref{eq:ratio}), which assumes that wind propagates along the equatorial plane and ignores thermal pressure ($c_{\rm s}=0$; \citealp{spruit96}). Our wind model has a more general application than the simplified model and differs from it by a constant angle from the equatorial plane and finite wind temperature. In \S\ref{sec:mu}, we will directly compare our results to the expressions derived from this simplified model.

The locus of the Alfv\'{e}n point can be expressed as 
\begin{equation}
\frac{R_{\rm A}}{R_0} = [\frac{3}{2}(1+\mu^{-2/3})]^{1/2}.
\label{eq:RA}
\end{equation}
As the wind is lightly loaded, the Alfv\'{e}n radius is far from the wind base. While for heavily loaded wind, the Alfv\'{e}n radius reaches a minimum of $R_{\rm A}/R_0=(3/2)^{1/2}$ when $\mu \to\infty$. Furthermore, the terminal wind velocity can be written by 
\begin{equation}
v_{\rm p}^{\infty}=\omega R_0\mu^{-1/3},
\end{equation}
which states that wind carrying small mass flux can be accelerated to large velocities. When $\mu=1$, the terminal velocity is equal to the rotational velocity at the wind base. 

The ratio of the toroidal to poloidal magnetic field at the Alfv\'{e}n radius can be approximated by
\begin{align}
\frac{B_\phi}{B_{\rm p}}\biggm|_{R_{\rm A}} & \approx (19/8)^{1/2} \;\; (\mu\ll 1), \nonumber \\
&\approx 1.14\mu \qquad\; (\mu\gg 1). 
\label{eq:ratio}
\end{align}
In the limit of weak mass loading, the ratio reaches a constant and is nearly unity. This case can be referred to as the centrifugally accelerated wind. Up to the Alfv\'{e}n radius, the field lines are not strongly bent and the wind corotates with the field line. On the other limit, with strong mass loading, the field line winds up developing strong toroidal magnetic field from the wind base with the corotation breaking down. The wind is then accelerated by the toroidal magnetic pressure gradient with terminal velocity much less than that at the wind base. 

Another quantity of interest is the ratio of Poynting flux to kinetic energy flux $\sigma$. The component of Poynting flux parallel to the poloidal field is $-\omega RB_{\rm p}B_\phi/4\pi$. Far beyond the Alfv\'{e}n radius $R\gg R_{\rm A}$, one expects $\Omega \approx \omega R_{\rm A}^2/R^2$ by Equation \eqref{eq:vphi} so that $\Omega \ll \omega$. Thereby, from Equation \eqref{eq:omega} one has $B_{\rm p} \approx B_\phi v_{\rm p}/\omega R$, hence $B_\phi \gg B_{\rm p}$ at very large $R$. Then, we can write the conversion of magnetic to kinetic energy by 
\begin{equation}
\sigma= \frac{-\omega RB_{\rm p}B_\phi}{2\pi\rho v_{\rm p}^3} \approx \frac{B_\phi^2}{2\pi\rho v_{\rm p}^2}\biggm|_\infty \approx 2\frac{v_{\rm A}^2}{v_{\rm p}^2} \biggm|_\infty.
\label{eq:sigma}
\end{equation}
Note that at infinity, we have $v_{\rm p} > v_{\rm A}$ since $v_{\rm fp} > v_{\rm A}$, which gives an asymptotic value of $\sigma<2$. 

\section{Model Parameters}\label{sec:bc}

The footpoint of the magnetized disk winds is set to be at a spherical radius $r_0=10^3r_{\rm g}$, where $r_{\rm g}\equiv GM/c^2$ is the gravitational radius. For convenience, we normalize radius, velocity, and density by their values at the footpoint of the field line, such that $R_0=v_{\rm K0}=\rho_0=1$. Poloidal magnetic field strength is parameterized by the poloidal Alfv\'{e}n velocity $v_{\rm Ap0} = B_{\rm p0}/\sqrt{4\pi\rho_0}$ at the wind base. This conveniently relates the magnetic field strength to the velocity so that one can describe the field strength by comparing it to the Keplerian velocity. For the fiducial wind model, the parameters are chosen to be
\begin{align}
&z_0=0.5R_0,  \qquad \theta=45^\circ, \nonumber \\
&v_{\rm Ap0}= 0.2v_{\rm K0}, \;\;\; c_{\rm s0}=0.5v_{\rm K0}, \qquad \omega=0.8\Omega_{\rm K0}, 
\label{eq:fid}
\end{align}
and the adiabatic index is set to $\gamma = 1.4$ throughout. 

The fiducial parameters are set by considering the regime of hot accretion flows. Wind properties are investigated in Y15 based on 3D GRMHD simulations where disk winds are produced self-consistently from hot accretion flows. These winds are launched from $\sim 30 r_g$ up to the outer boundary of the accretion flow, which implies that at the footpoint winds are a combination of those originated from $r\leq r_0$. Launched from different radii, winds possess different velocities and almost keep constant during the outward propagation, indicating that the velocity at the footpoint must be diverse. The trajectory of wind after launching is found to follow a straight line along an angle $\theta \lesssim 45^\circ$, i.e. more prone to the pole rather than the equatorial plane. 
 
\begin{figure*}[ht!]
\epsscale{1.2}
\plotone{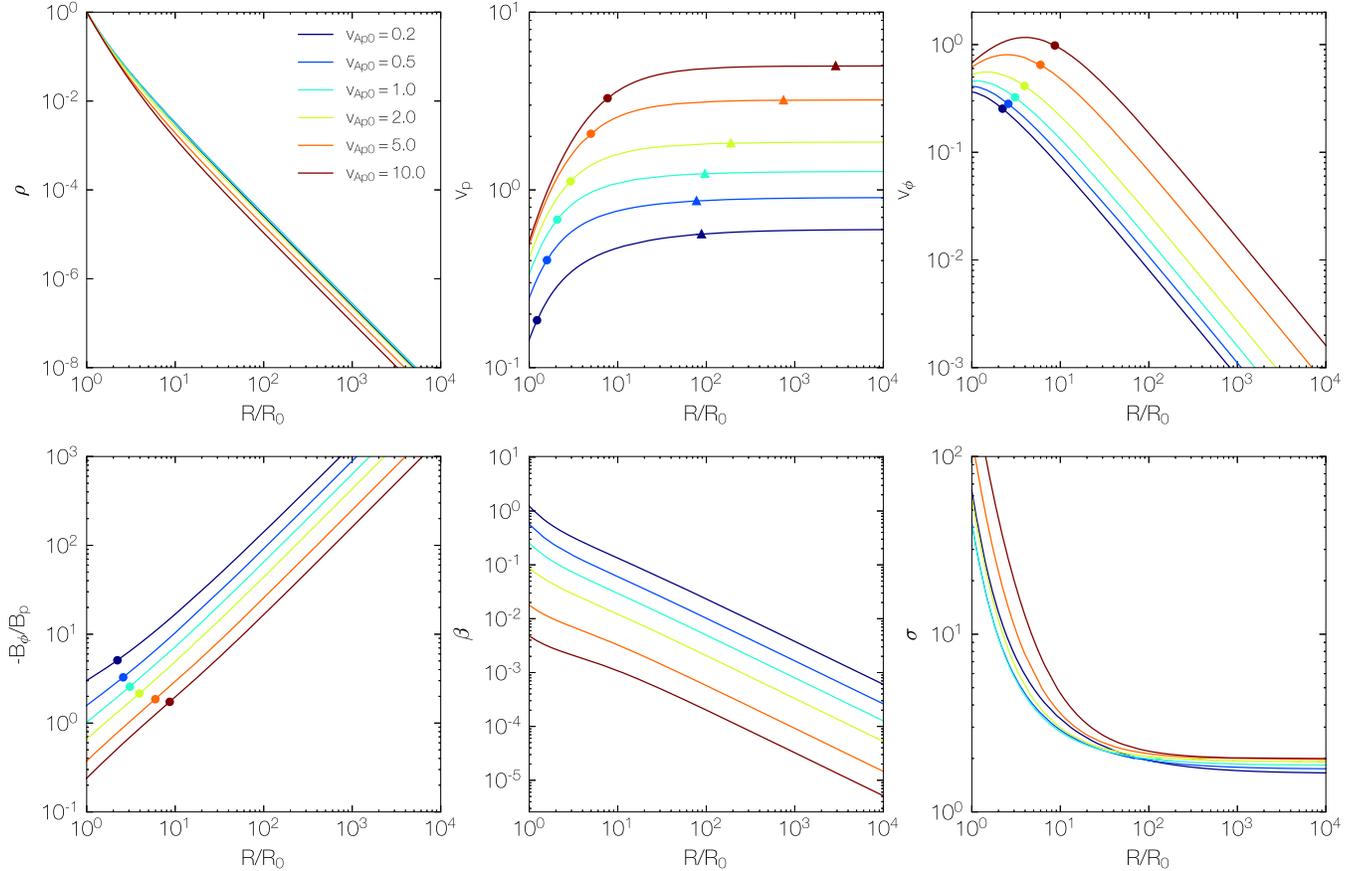}
\caption{Profiles of density, poloidal and toroidal velocity, the ratio of toroidal to poloidal magnetic field, plasma $\beta$, and the ratio of Poynting flux to kinetic energy flux $\sigma$ of wind as a function of cylindrical radius $R$ along the wind trajectory prescribed with $\theta=45^\circ$ to the rotational axis. Colors indicate various initial poloidal magnetic field strengths with $v_{\rm Ap0}$ ranging from 0.2 to 10 $v_{\rm K0}$. Solid circles mark the Alfv\'{e}n points, and triangles represent the fast magneto-sonic points.}
\label{fig:fid}
\end{figure*}

We adopt wind properties concluded in Y15 as our fiducial model parameters. The location of the footpoint is chosen to be the outer radial boundary of their simulations. The hot accretion flow maintains a disk aspect ratio which the scale height $H$ is about a half of the cylindrical radius $R$. We assume the wind base is at one disk scale height and set $\theta = 45^\circ$ in the fiducial model. The poloidal magnetic field strength $v_{\rm Ap0}$ is adopted so that the plasma $\beta \equiv P_{\rm gas}/P_{\rm mag}=8\pi\rho c_{\rm s}^2/(B_{\rm p}^2+B_\phi^2)$, defined as gas pressure over magnetic pressure, is about unity at the wind base (Figure \ref{fig:fid}). The sound speed is set by disk aspect ratio $H/R  = c_{\rm s0}/v_{\rm K0}$, which is equivalent to about $1.36\times10^9$ K at the footpoint. The fiducial angular velocity of the field line $\omega$ is computed by Equation \eqref{eq:omega} in accretion disk simulations. We address that the poloidal velocity at the footpoint is not prescribed in Equation \eqref{eq:fid}, different from our previous hydrodynamic work, because satisfying conditions of passing through all three critical points smoothly places constraints on the number of parameters needed to be given. Hence, the poloidal velocity $v_{\rm p0}$ is solved by MHD equations, and we confirm that the value found in the fiducial setup is consistent with the value of GRMHD simulations (see \S\ref{sec:vap0}).     

Winds emerged from thin disks have different properties from hot accretion flows. The disks are cold for which the sound speed $c_{\rm s0}$ is expected to be low. A value of 0.1 or 0.05$v_{\rm K0}$ is usually taken for these disks in numerical simulations. For both accretion regimes, the magnetic field strengths at the wind base are barely constrained. Aiming to include a variety of winds with diverse properties from hot accretion flows and thin disks, we employ parameter spaces as follows besides our fiducial setup. We obtain wind solutions over large domains of poloidal magnetic field, where $v_{\rm Ap0} \in [0.01,100]$. The temperatures at the wind base span over $c_{\rm s0} \in [0.01,0.5]$. The angular velocity of the field line $\omega \in \{0.8, 1\}$ and the adiabatic index $\gamma \in \{1.3,1.4,1.5\}$ are also under investigation in \S\ref{sec:parastudy}. 

\section{Results}\label{sec:result}

In this section, we present numerical results of MHD wind solutions by analyzing the fiducial model in \S\ref{sec:fid}, the dependence on poloidal magnetic field strength in \S\ref{sec:vap0}, the wind temperature in \S\ref{sec:cs0}, and the mass loading in \S\ref{sec:mu}. The wind acceleration mechanism is discussed in \S\ref{sec:acc}. We address that all the solutions obtained have sub-Alfv\'{e}nic velocities at the wind base, though some of the solutions may already pass through the slow magneto-sonic point. 

\subsection{The Fiducial Solution}\label{sec:fid}

We commence with analyzing the behaviour of characteristic physical quantities with fiducial parameters prescribed in Equation \eqref{eq:fid}, showing by curves of $v_{\rm Ap0}=0.2$ in Figure \ref{fig:fid}. The radial profiles of density, poloidal and toroidal velocities, ratio of magnetic field, plasma $\beta$, and ratio of Poynting to kinetic energy flux are displayed in the plot. 

The upper middle panel indicates that the wind is accelerated monotonically passing through Alfv\'{e}n and fast magneto-sonic points. We address that the poloidal velocities of the fiducial model already passed through the slow magneto-sonic point at the wind base. The wind keeps accelerating after propagating through the Alfv\'{e}n point. The poloidal velocity approaches an asymptotic value at large radii; beyond fast point, it almost maintains a constant. From the figure, the density profile drops as $\rho \propto R^{-2}$ at large distances which is as expected in Equation \eqref{eq:kappa}, since the poloidal magnetic field strength is prescribed to obey the divergence free condition as $B_{\rm p}\propto R^{-2}$, and the poloidal velocity keeps about a constant beyond the fast point. 

The angular velocity profile shown in the upper right panel implies whether the gas corotates with the field line, i.e.  $v_{\phi} \propto R$, or it rotates by conserving its specific angular momentum, i.e. $v_{\phi} \propto R^{-1}$. These two regimes correspond to different wind acceleration mechanisms which will be detailed in \S\ref{sec:acc}. In short, when the poloidal magnetic field dominates the toroidal component, the corotation occurs and associates with the magneto-centrifugal force. During the outward propagation of the gas, the field lines wind up with the development of toroidal magnetic fields. The corotation is ceased once the toroidal component dominates. The wind then rotates by conserving specific angular momentum, and the acceleration is driven by the toroidal magnetic pressure gradient. In our fiducial model, the gas is mainly in the toroidal magnetic pressure gradient driven case.

In the lower left panel, we display the ratio of toroidal to poloidal magnetic field. A minus sign is taken since the toroidal magnetic field has opposite sign to the poloidal one both above and below the equatorial plane due to the disk rotation. Since $-B_\phi/B_{\rm p}$ scales approximately as $\propto R$, the toroidal field strength possesses a flatter slope than the poloidal one with $B_\phi \propto R^{-1}$. The plasma $\beta$ is computed by the ratio of gas pressure to magnetic pressure. Although $B_{\rm p}$ and $B_\phi$ are comparable at the wind base, the magnetic pressure is dominated by toroidal component at large radii. The plasma $\beta$ is then dominated by the profile of sound speed as the density and the magnetic pressure have the same proportionality with $R$. The sound speed is proportional to $c_{\rm s}^2 \propto \rho^{\gamma-1}$ so that one can obtain $\beta \propto c_{\rm s}^2 \propto R^{-0.8}$. 

The conversion of Poynting flux to kinetic energy flux are shown in the lower right panel of Figure \ref{fig:fid}. Near the wind base, the magnetic energy overwhelms the kinetic energy. As wind propagating outward, the magnetic energy converts to kinetic energy yielding a decline in their ratio. Beyond the fast magneto-sonic point, the ratio approaches an asymptotic value of $\sigma\sim2$ as expected in Equation \eqref{eq:sigma}. 

\subsection{Dependence on Poloidal Magnetic Field}\label{sec:vap0}

\begin{figure*}[ht!]
\epsscale{1.2}
\plotone{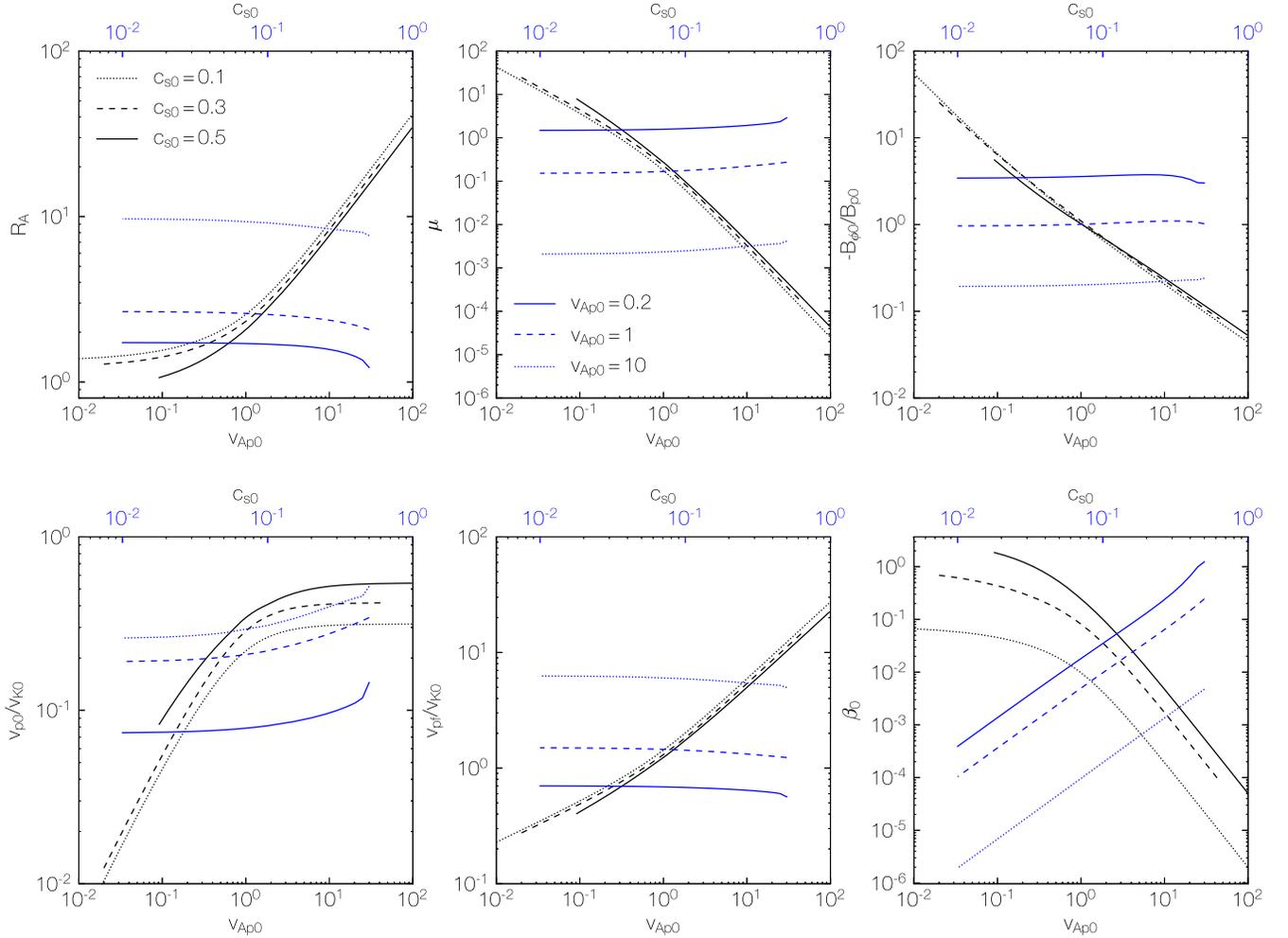}
\caption{Alfv\'{e}n radius, mass loading parameter, ratio of toroidal to poloidal magnetic field strength, poloidal velocity at the wind base and at fast magneto-sonic point, and plasma $\beta$ at wind base as a function of poloidal magnetic field $v_{\rm Ap0}$ (black curves) and temperature $c_{\rm s0}$ (blue curves) at wind base. When varying $v_{\rm Ap0}$, three values of temperature are investigated where $c_{\rm s0} = 0.1$ (dotted), 0.3 (dashed), and 0.5 (solid). When varying $c_{\rm s0}$, three values of poloidal field $v_{\rm Ap0}$ are investigated where $v_{\rm Ap0} = 0.2$ (solid), 1 (dashed), and 10 (dotted). }
\label{fig:cs}
\end{figure*}

To study the dependence of poloidal magnetic fields, we keep $c_{\rm s0}$ constant and vary $v_{\rm Ap0}$. Since the initial poloidal magnetic field strength is barely constrained, we explore a large domain by setting $v_{\rm Ap0}=0.01v_{\rm K0}$ and $v_{\rm Ap0}=100v_{\rm K0}$ to be the lower and upper limit (Figure \ref{fig:cs}). The lower values of $v_{\rm Ap0}$ (weak magnetic field strength) can be associated to the standard and normal evolution (SANE; \citealp{narayan_etal12}) model referred in hot accretion flow simulations, and larger values of $v_{\rm Ap0}$ (strong magnetic field strength) can be related to the magnetically arrested disk (MAD; \citealp{narayan03}) model.

In Figure \ref{fig:fid}, we show profiles of diagnostic physical quantities at various $v_{\rm Ap0}$ with non-consecutive values from $0.2v_{\rm K0}$ to $10v_{\rm K0}$. The overall proportionality as a function of $R$ of each physical quantity at large distances for different $v_{\rm Ap0}$ shares great similarity. Larger poloidal field strengths lead to fast poloidal velocities. The Alfv\'{e}n points and fast magneto-sonic points generally shift toward larger radii as poloidal fields are enhanced. The poloidal velocities all tend to approach an asymptotic value at large radii. At fast magneto-sonic point, the poloidal velocity mostly reaches its asymptotic value. The angular velocity in Figure \ref{fig:fid} shows a transition from corotation to conserving specific angular momentum for strong poloidal fields. When the poloidal magnetic field is weak at the wind base, the magnetic tension is not able to sustain the corotation between the gas and the field line (see \S\ref{sec:acc}). The ratio of toroidal to poloidal magnetic field tends to be smaller once the poloidal field strength is stronger at the launching point as expected, and the plasma $\beta$ drops with increasing $v_{\rm Ap0}$.

In Figure \ref{fig:cs}, we show in black curves the dependence of poloidal field strength $v_{\rm Ap0}$ on Alfv\'{e}n point, mass loading parameter, magnetic field strength ratio at wind base, poloidal velocity at the launching point and at fast magneto-sonic point, and the plasma $\beta$ at wind base. The filed strength spans over 0.01 to 100 $v_{\rm K0}$. Shown in the upper left panel, the Alfv\'{e}n point shifts toward large radii with $v_{\rm Ap0}$ and boosts when $v_{\rm Ap0} \gtrsim 1$. The ejection index (Equation \ref{eq:xi}) which is the ratio of cumulative mass-loss rate to wind-driven mass accretion rate is directly related to the location of the Alfv\'{e}n point. Taking representative values of $v_{\rm Ap0}=0.1, 1$, and $10$, we find the corresponding ejection indices $\xi \approx 3, 0.15$ and $0.0086$, respectively. Larger ejection indices correspond to more massive mass loading. As shown in the upper middle panel, the mass loading factor is a decreasing function with $v_{\rm Ap0}$, hence an increasing function of $\xi$ as expected. The field strength ratio $|B_{\phi0}/B_{\rm p0}|$ declines with $v_{\rm Ap0}$, which indicates strong mass loading leads to fast development of toroidal magnetic field since it is harder to enforce corotation with more massive winds so that the field line bends more.  

The lower left panel of Figure \ref{fig:cs} delineates poloidal velocity at the wind base. In the domain of $v_{\rm Ap0}$ we investigated, the poloidal velocity at the footpoint $v_{\rm p0}$ spans over 0.01 to 0.5 $v_{\rm K0}$ with increasing poloidal field strength. Small-scale 3D GRMHD simulations of hot accretion flows imply that near the surface a poloidal velocity at each radius is about 0.2$v_{\rm K}(r)$, where $v_{\rm K}(r)$ is the local Keplerian velocity (Y15). Their results are achieved via weighting through the mass flux of wind at different radii, since at each launching point $r_0$ the outflow is a combination of wind originated from $r<r_0$. One can  approximately treat $0.2v_{\rm K}$ to be the wind poloidal velocity launched at $r_0$ since the wind launched at larger radii carries more mass flux, concluded by fitting the simulation results, which gives $\dot M_\mathrm{wind} \propto r^s$ and $s\approx 1$. In \S\ref{sec:bc}, we address that the poloidal velocity at footpoint is not prescribed because the number of parameters to be given at the boundary are limited by requiring wind solutions to pass through the slow and fast critical points. With our fiducial wind temperature $c_{\rm s0}=0.5$ and fiducial field strength at wind base $v_{\rm Ap0} =0.2$, the poloidal velocity is $v_{\rm p0} \approx 0.2v_{\rm K0}$, which is consistent with the GRMHD simulations. We also note that when $v_{\rm Ap0} \gtrsim 1$, the poloidal velocity at wind base approaches an asymptotic value of about 0.5$v_{\rm K0}$.

The terminal velocity of wind also deserves attention. It reaches a faster speed as poloidal magnetic field at the wind base goes larger. The poloidal velocity can rise several times or one order-of-magnitude larger than it is at the wind base. The fiducial model of $v_{\rm Ap0}=0.2$ reveals a terminal velocity of  $v_{\rm pf} \approx 0.5v_{\rm K0} (\approx 0.016c)$. Y15 trace the trajectory of wind from $80$ to $10^3 r_{\rm g}$ to study the physical properties during the outward propagation of wind. They find the poloidal velocity of wind with opening angles $\theta \lesssim 30^\circ$ shows an increase with distance, and it tends to keep constant since the launching point when $40^\circ \lesssim \theta \lesssim 50^\circ$. Their results are applicable to a radial extent close to the accretion disk (up to a few times 100$r_{\rm g}$) where the corona region above the main disk body is rather turbulent, unlike the pure MHD model adopted in this work.

It is immediately apparent that the plasma $\beta_0$ is a decreasing function with $v_{\rm Ap0}$. As seen in the lower right panel of Figure \ref{fig:cs}, it ranges from unity down to $10^{-4}$ for $v_{\rm Ap0}\in[0.01,100]$ for the fiducial temperature $c_{\rm s0}=0.5$, and even lower when the temperature drops down. 
Another diagnostic quantity of interest is the ratio of Poynting flux to kinetic energy flux $\sigma$. Though not shown in the figure, we address that the Poynting flux to kinetic energy flux ratio is nearly a constant $\sigma \sim 2$ at large radii through the entire domain of $v_{\rm Ap0}$ and $c_{\rm s0}$ employed in Figure \ref{fig:cs}. This means the values of Alfv\'{e}n velocity and fast magneto-sonic velocity are comparable toward large radii (Equation \ref{eq:sigma}).

\subsection{Dependence on Temperature}\label{sec:cs0}

We study the influence of wind temperature at launching point since it differs substantially for hot accretion flows and thin disks. In Figure \ref{fig:cs}, we first keep $c_{\rm s0}$ fixed throughout the domain of $v_{\rm Ap0}$ by adopting three representative values $c_{\rm s0}=0.1,0.3$, and 0.5 $v_{\rm K0}$. Then we vary $c_{\rm s0}$ at fixed values of $v_{\rm Ap0}=0.1,1$, and 10 $v_{\rm K0}$. We find that wind properties do not show strong dependency on the temperature at the footpoint, except plasma $\beta_0$. 

The black curves in Figure \ref{fig:cs} reveals that the influence of launching point temperature on wind evolution is modest. The enhanced wind temperature results in smaller Alfv\'{e}n radius, higher mass loading, faster launching velocity, and slower terminal velocity. The magnetic field shows equal partition between poloidal and toroidal components for all three $c_{\rm s0}$ at $v_{\rm Ap0} \sim 1$. With slopes of $|B_{\phi0}/B_{\rm p0}|$ varying slightly with different $c_{\rm s0}$, the wind at footpoint is generally more toroidal field dominated when $v_{\rm Ap0} < 1$ and vice versa. When we fix $v_{\rm Ap0}$ at $0.2, 1$, and 10 $v_{\rm K0}$, wind temperature $c_{\rm s0}$ from $0.01$ to 0.5 $v_{\rm K0}$ are under investigation. Shown in the blue curves in Figure \ref{fig:cs}, the diagnostic quantities generally vary slightly with $c_{\rm s0}$, except $v_{\rm Ap0}$ and $\beta_0$. The poloidal field  $v_{\rm Ap0}$ shows more prominent increase with $v_{\rm Ap0} \gtrsim 1$. The plasma $\beta_0$ at wind base is a strong function of $c_{\rm s0}$ since it is closely related to the gas pressure. The black and blue curves in the lower right panel of Figure \ref{fig:cs} indicates that at fixed $v_{\rm Ap0}$ higher temperature yields lower $\beta_0$ as expected.

\subsection{Acceleration Mechanism}\label{sec:acc}

\begin{figure*}[ht!]
\epsscale{1.2}
\plotone{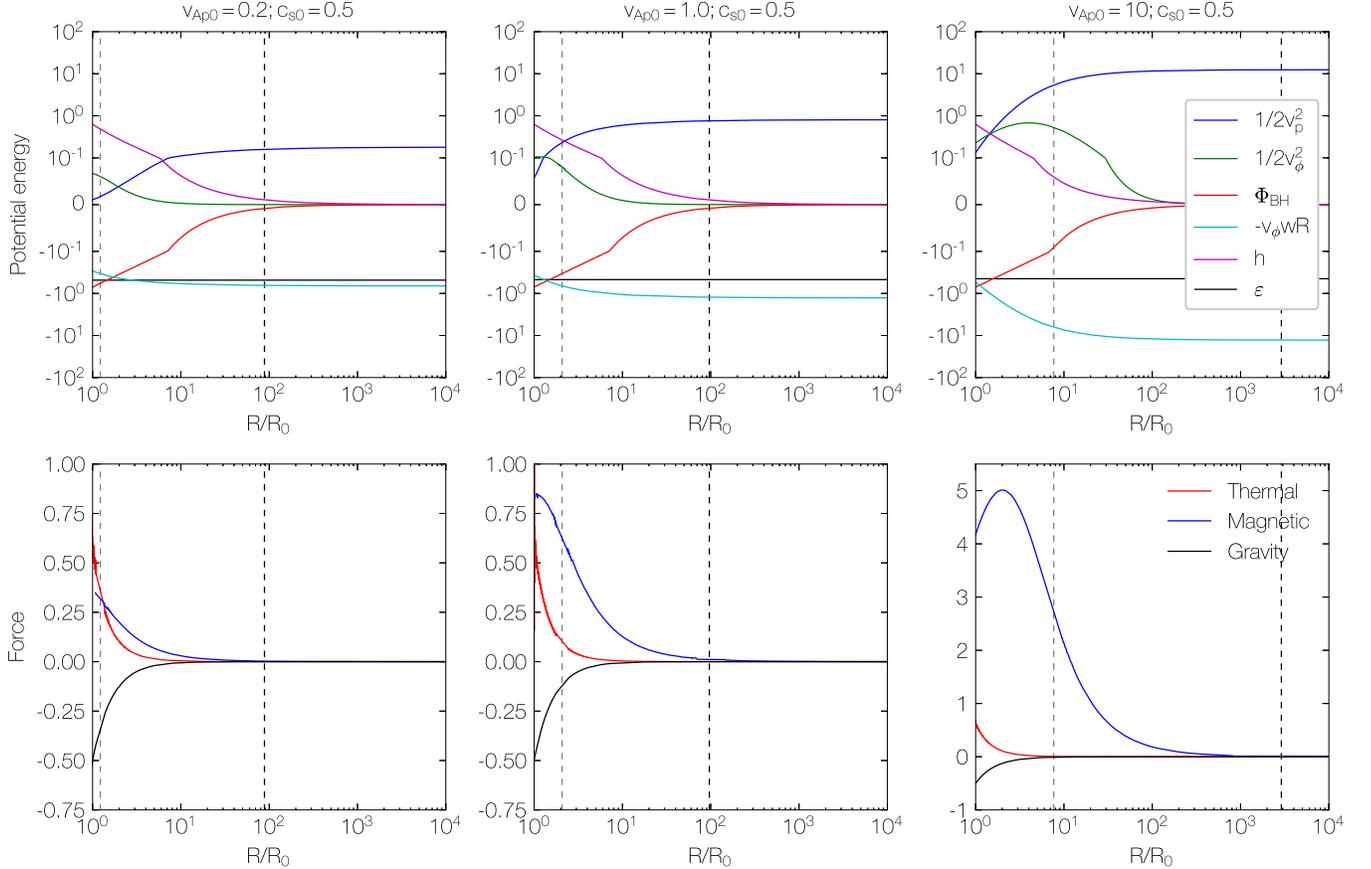}
\caption{Components of potential energies and forces along the wind trajectory with varying $v_{\rm Ap0}=0.2$ (first column), $1$ (second column), $10$ (third column) and fixed  $c_{\rm s0}=0.5$. Top panels: Bernoulli constant ($\varepsilon$) and components of potential energies, namely, radial kinetic energy ($1/2v^2_\mathrm{p}$), rotational energy ($1/2v^2_\phi$), black hole gravitational potential ($\Phi$), enthalpy ($h$), and centrifugal potential ($-v_\phi \omega R$) as a function of $R$ (Equation \ref{eq:e1}). Bottom panels: components of forces, namely, thermal pressure gradient (red), magnetic pressure gradient (blue), and gravity (black) as a function of $R$ (Equation \ref{eq:force}). The vertical dashed lines denote the Alfv\'{e}n points (grey) and fast magneto-sonic points (black).}
\label{fig:acc}
\end{figure*}

In this work, we aim to study how magnetism influences the dynamics of disk winds. The magnetically driving mechanism can be divided into two categories based upon the locus of the Alfv\'{e}n point (Equations \ref{eq:fieldgas1} and \ref{eq:fieldgas2}). In general, with radii smaller than the Alfv\'{e}n radius, the gas is accelerated mainly via magneto-centrifugal force where corotation is enforced. Beyond the Alfv\'{e}n radius, gas conserves specific angular momentum and is accelerated via toroidal magnetic pressure gradient.

Physically, near the wind base the poloidal magnetic field strength is reasonably large so that the magnetic pressure is strong compared to the gas pressure or the ram pressure. The magnetic tension force persists and the field line behaves like a rigid wire where gas is free to move along it, resembling the scenario of ``beads on a wire''. The wind is enforced to corotate with the field line sticking out of the disk surface. The enforced corotation causes the increase of centrifugal force with distance to sustain the outward acceleration. This regime is referred to magneto-cetrifugal force in driving outflows. Along the wind trajectory, the poloidal field strength drops toward large distances. The magneto-centrifugal acceleration will effectively stop when the ram pressure starts to exceed magnetic pressure, and the corotation is ceased to be valid since the magnetic tension force weakens. Meanwhile, the toroidal component of the field builds up due to the disk rotation and subsequently dominates over its poloidal component. Then the flow is accelerated mainly through toroidal magnetic pressure gradient.  

In Figure \ref{fig:fid}, the upper right panel shows that with strong poloidal magnetic field at the wind base ($v_{\rm Ap0}\gtrsim5$), the corotation is enforced near the disk surface with poloidal magnetic field dominates the toroidal component as seen in the lower left panel. Close to the Alfv\'{e}n radius, $|B_\phi/B_{\rm p}|$ becomes above unity, and the corotation is ceased. Weak poloidal magnetic fields ($v_{\rm Ap0}\lesssim5$) possess a large $|B_\phi/B_{\rm p}|$ at the wind base, not even allowing the corotation to occur.

A more intuitive understanding of wind acceleration mechanism can be achieved through looking at the components of specific energy terms in the Bernoulli integral (Equation \ref{eq:e}), which is expressed in a frame rotating with angular frequency $\omega$. Rearranging Equation \eqref{eq:e}, we can arrive at 
\begin{align}
\varepsilon = \frac{1}{2}v^2_\mathrm{p}+\frac{1}{2}v_\mathrm{\phi}^2+\Phi+h-v_\phi \omega R.
\label{eq:e1}
\end{align}
On the right hand side of Equation \eqref{eq:e1}, the terms correspond to radial kinetic energy, rotational energy, gravitational potential, enthalpy, and centrifugal potential, respectively. In the upper panels of Figure \ref{fig:acc}, we show components of Bernoulli integral as a function of $R$ at $v_{\rm Ap0}=0.2, 1$, and 10$v_{\rm K0}$. In the limit of weak poloidal field ($v_{\rm Ap0}=0.2 v_{\rm K0}$), it suggests that the drop of enthalpy, rotational energy, and centrifugal potential compensates the increase of the gravitational potential and radial kinetic energy. More precisely, it is mainly the rapid decrease of enthalpy that offsets the quick growth of the gravitational potential, which is consistent with the results in \citet{cyl19} where pure hydrodynamic model is assumed. Towards strong poloidal field limit ($v_{\rm Ap0}=10 v_{\rm K0}$), the radial kinetic energy shows a more pronounced increase primarily due to the energy converted from the centrifugal potential. Intermediate poloidal field ($v_{\rm Ap0}=1$) leads to a case in between.

It is noteworthy that the Bernoulli integral (Equation \ref{eq:e}) has no contribution from magnetic forces as in the corotating frame the total magnetic field ${\bold B}$ is parallel to the total velocity ${\bold v}$. Nonetheless, it is ultimately the magnetic forces that drive the outward propagation of the wind. One can find that the magnetic term does involve in the Bernoulli integral in its rest frame expression (\S\ref{sec:eqn}). To examine the effects of magnetism on driving disk winds, we write the equation of motion along the poloidal magnetic field as 
\begin{align}
\frac{dv_{\rm p}}{dt} = -\frac{1}{\rho}\frac{dP}{ds} - \frac{d\Phi}{ds} - \frac{1}{8\pi\rho}\frac{d B_\phi^2}{ds},
\label{eq:force}
\end{align}
where $s$ is the length along the wind trajectory. The last term on the right hand side associates to the pressure gradient of toroidal fields along the direction of wind propagation. The outward acceleration of wind requires the thermal and magnetic pressure gradient to overcome the gravity. 

In the bottom panels of Figure \ref{fig:acc}, we decompose the poloidal forces into thermal and magnetic pressure gradient, as well as gravity at different $v_{\rm Ap0}$ and fixed $c_{\rm s0}=0.5v_{\rm K0}$. In the limit of weak poloidal field ($v_{\rm Ap0}=0.2 v_{\rm K0}$), the forces exerted by thermal and magnetic pressure gradient are comparable, with thermal pressure gradient being more pronounced before passing through the Alfv\'{e}n point. The toroidal magnetic pressure gradient dominates over the thermal pressure to drive outward acceleration when $v_{\rm Ap0}>1$. The magnetic pressure gradient becomes overwhelming when the poloidal field at the wind  base is strong, i.e. $v_{\rm Ap0}\gtrsim 10$, which results in the boost of terminal velocity (Figure \ref{fig:cs}).

\subsection{Dependence on Mass Loading}\label{sec:mu}

\begin{figure}[t!]
\epsscale{1.25}
\plotone{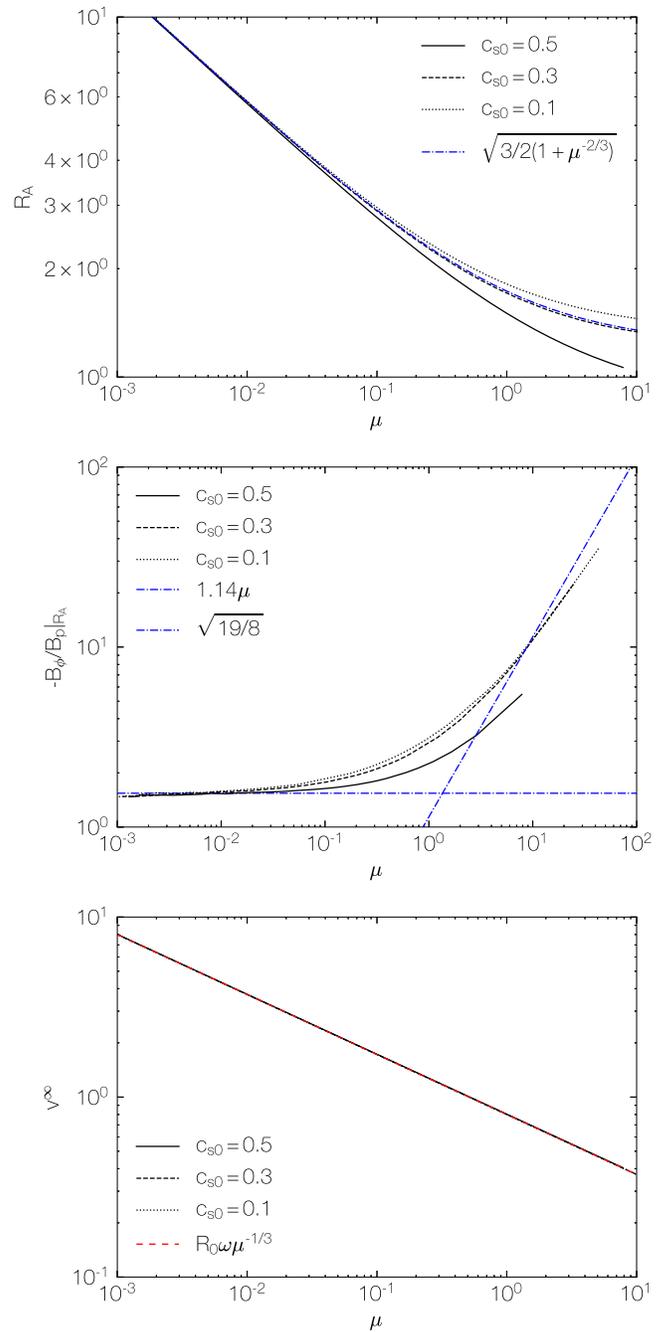}
\caption{Diagnostic physical quantities as a function of mass loading parameter $\mu$ at various wind base temperatures $c_{\rm s0}=0.5$ (solid black), 0.3 (dashed black), 0.1 (dotted black). Top panel: loci of Alfv\'{e}n point as a function of mass loading parameter. Middle panel: ratio of poloidal to toroidal magnetic field strength at Alfv\'{e}n point. Bottom panel: wind terminal velocity as a function of $\mu$. The dash-dotted blue curves in all panels denote the relation derived from cold Weber \& Davis wind model in \S\ref{sec:pred}.}
\label{fig:mu}
\end{figure}

In \S\ref{sec:pred}, we present (asymptotic) relations between diagnostic physical quantities (Equations \ref{eq:RA}-\ref{eq:sigma}), namely, the Alfv\'{e}n radius, the terminal velocity, the ratio of magnetic field strength, the ratio of Poynting flux to kinetic energy flux, for which Equations \eqref{eq:RA}-\eqref{eq:ratio} are derived in the case of cold Weber \& Davis wind model ($\theta = 90^\circ, c_{\rm s0}=0$). Our model differs from it by an inclined wind trajectory $\theta = 45^\circ$ and $c_{\rm s0}=0.1,0.3$, and $0.5$. Being more generalized, our results are compared to the derived relations to test whether they can still be obeyed.

In Figure \ref{fig:mu}, the top panel shows that high mass loading (weak $v_{\rm Ap0}$) is associated to small Alfv\'{e}n radius so large mass loss to mass accretion rate. Our results match perfectly to the relation in Equation \eqref{eq:RA} especially when $\mu \lesssim 0.1$ for all three $c_{\rm s0}$ adopted. It seems the relation holds for small mass loading. When $\mu \gtrsim 0.1$, winds with different $c_{\rm s0}$ show deviations from the derived relation in different extents. The $R_{\rm A}$ falls below expectation for $c_{\rm s0}= 0.3$ and $0.5$, but rises up for $c_{\rm s0}= 0.1$. A lower limit of $R_{\rm A}=(3/2)^{1/2}$ is placed for high mass laoding by the cold wind model, while it is no longer valid for winds that possess finite temperature.

The middle panel of Figure \ref{fig:mu} implies that high mass loading results in more toroidal magnetic field dominated case, since it is more difficult to enforce gas to corotate with the field line once the outflow is massive. The horizontal asymptotic relation is strictly obeyed when $\mu \ll 1$ (strong $v_{\rm Ap0}$) for all $c_{\rm s0}$. In the weak poloidal field limit, colder winds show better consistency to the derived relation, while the curve of $c_{\rm s0}=0.5$ deviates more prominently from the asymptotic line. Given $\mu$ fixed, warmer winds tend to obtain lower $|B_\phi/B_{\rm p}|_{\rm R_A}$ values, because that the Alfv\'{e}n radius is closer to the wind base and that $|B_\phi/B_{\rm p}|_{\rm R_A}$ is always an increasing function with radius.

In the bottom panel, we show the terminal velocity as a function of mass loading $\mu$. Despite of the wind base temperature, all three models with various $c_{\rm s0}$ are strictly satisfied the asymptotic relation. It is likely because that our prescribed wind model, with an adiabatic index $\gamma=1.4$, is cooled nearly adiabatically. Hence, the wind temperature drops rapidly with radius. At large distances, the wind is cold as in the model for the derived relation. 

\section{Parameter Study} \label{sec:parastudy}

\begin{figure*}[ht!]
\epsscale{1.2}
\plotone{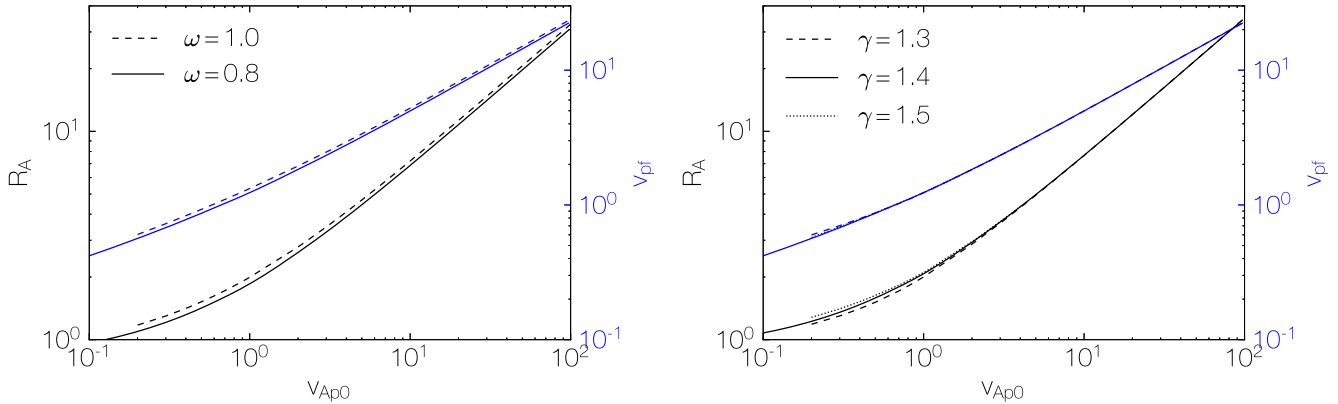}
\caption{ Alfv\'{e}n radius $R_{\rm A}$ (black) and terminal velocity $v_{\rm pf}$ (blue) as a function of poloidal field strength $v_{\rm Ap0}$ at various field line angular velocities $\omega\in\{0.8,1\}$ and adiabatic indices $\gamma\in\{1.3,1.4,1.5\}$.}
\label{fig:para}
\end{figure*}

We instigate the dependance of diagnostic physical quantities on field line angular velocity (\S\ref{sec:fieldangv}) and adiabatic index (\S\ref{sec:adindex}) in this section. In Figure \ref{fig:para}, we show two quantities of interest, namely $R_{\rm A}$ and $v_{\rm pf}$, as a function of poloidal magnetic field strength at the footpoint $v_{\rm Ap0}$. The Alfv\'{e}n point $R_{\rm A}$ directly relates to the mass loading and can be used to discriminate the magneto-centrifugal and toroidal magnetic pressure gradient dominated region. The terminal velocity $v_{\rm pf}$ shows the extent of wind acceleration. 

\subsection{Dependence on Field Angular Velocity}\label{sec:fieldangv} 
     
We compute the angular velocity of the field line using Equation \eqref{eq:omega} with small-scale accretion disk simulations. The $\theta$-dependence leads us to adopt a larger value of $\omega=1$ to follow the trend of solutions. In the lower left panel of Figure \ref{fig:para}, we find that higher field angular velocity results in a greater $R_{\rm A}$ and $v_{\rm pf}$.

\subsection{Dependence on Adiabatic Index}\label{sec:adindex} 

The adiabatic index $\gamma$ of wind remains uncertain and is determined by the intricate interplay among thermal conduction, heating and cooling. Accurate values can be obtained through numerical simulations by following the trajectory of wind which may introduce further complications to our MHD wind equations by varying its value along the field line. In the hydrodynamic case, $\gamma<1.5$ is required to obtain transonic solutions when angular momentum is not taken into account. 

For simplicity, we adopt a constant $\gamma$ in the model and test adiabatic indices above and below our fiducial value. In the right panel of Figure \ref{fig:para}, the wind solution is generally not a strong function of $\gamma$, provided by the fact that $R_{\rm A}$ and $v_{\rm pf}$ do not show discernible variations when $v_{\rm Ap0}$ is greater than unity. When $v_{\rm Ap0}$ is less than unity, larger adiabatic indices result in greater Alfv\'{e}n radii. That is caused by the thermal pressure gradient being comparable to the magnetic force in driving winds at weak field strengths. The temperature which is determined by $\gamma$ then plays an important role. As $v_{\rm Ap0}>1$, the magnetic force dominates the wind acceleration (see Figure \ref{fig:acc}) so that slight variation in $\gamma$ will not modify the solutions in a great extent. 

\section{Conclusions and Discussion} \label{sec:cd}

\subsection{Summary} \label{sec:sum}

In this work, we present an initial effort toward studying the dynamics of black hole accretion disk winds toward large radii. Disk winds are essential ingredients for AGN feedback in understanding the coevolution between the central supermassive black hole and the host galaxy. The limited dynamical range of small-scale accretion disk simulations does not allow us to study the kinematics of winds toward galaxy scales. In this work, we employ wind properties obtained in small-scale accretion disk simulations as our inner boundary conditions and adopt analytic model to provide a simple but intuitive way to understand wind dynamics over a wider spatial range.

We construct 1D MHD equations following \citet{wd67} in cylindrical coordinates. Four equations associated to four conserved quantities, including mass to magnetic flux, angular velocity of the field line, specific angular momentum, and specific energy (Equations \ref{eq:kappa}-\ref{eq:e}) are solved. The solution is requested to pass through the slow, Alfv\'{e}n, and fast critical points smoothly. We do not impose a condition that all three critical points should have their loci beyond the wind base. Our fiducial model is set with parameters for winds from hot accretion flows, specifically $c_{\rm s0}/v_{\rm K}=0.5, v_{\rm Ap0}/v_{\rm K}=0.2$, and $\omega/\Omega_{\rm K}=0.8$. The geometry of poloidal magnetic field is prescribed as a straight line with a constant angle from the rotational axis ($\theta=45^\circ$), while the strength is described by the divergence free condition. We summarize our main results as follows.

The physical quantities possess the following relations with cylindrical radius $R$ as the wind passes the fast magneto-sonic point:
\begin{align}
&\rho \propto R^{-2}  \qquad \;\;\; v_{\rm p}\propto {\rm const.} \qquad v_{\rm \phi}\propto R^{-1} \nonumber \\
&B_{\rm \phi}\propto R^{-1}  \qquad \beta \propto \rho^{\gamma-1},
\label{eq:propto}
\end{align}
with the prescribed poloidal magnetic field $B_{\rm p}\propto R^{-2}$. Moreover, we explore the dependence of poloidal magnetic field at wind base characterized by $v_{\rm Ap0}$ in a range from 0.01 to 100. The weak magnetic field case corresponds to SANE model in accretion flow simulations, and the strong magnetic field case associates to MAD model. The Alfv\'{e}n radius is a quick increasing function with magnetization when $v_{\rm Ap0}$ is above unity, whereas the mass loading parameter is a decreasing function of $v_{\rm Ap0}$. Equal partition of $|B_{\phi 0}/B_{\rm p0}|$ is achieved when $v_{\rm Ap0}$ is about unity, with smaller $|B_{\phi 0}/B_{\rm p0}|$ toward large $v_{\rm Ap0}$. The poloidal velocity at footpoint $v_{\rm p0}$ is enlarged with $v_{\rm Ap0}$ but approaches an asymptotic value of $0.5v_{\rm K0}$ when $v_{\rm Ap0}>1$. Faster terminal velocity $v_{\rm pf}$ associates to stronger $v_{\rm Ap0}$, and the plasma $\beta_0$ is a decreasing function of magnetization as expected. We further investigate the dependence of temperature at wind base $c_{\rm s0}$ from 0.01 to 0.5$v_{\rm K0}$, which shows modest impacts on physical quantities.

The wind acceleration mechanism is studied under different poloidal magnetic field strengths at the wind base. With strong poloidal fields $v_{\rm Ap0} \gtrsim 5$, the corotation can be enforced close to the disk surface. Beyond the Alfv\'{e}n radius, corotation is ceased where the gas rotates by conserving specific angular momentum. Weak poloidal fields $v_{\rm Ap0} \lesssim 5$ do not give rise to the corotation. The decomposition of Bernoulli constant reveals that with weak poloidal field where $v_{\rm Ap0}=0.2$, it is chiefly the enthalpy that converts to the black hole potential energy, resembling the scenario in pure hydrodynamic model. Strong poloidal field strength $v_{\rm Ap0}=10$ results in fast rise of radial kinetic energy attributed to the conversion from the centrifugal potential. The decomposition of poloidal forces indicate that the thermal pressure gradient is comparable to the toroidal magnetic pressure gradient near the wind base at $v_{\rm Ap0} = 0.2$, and the magnetic force dominates the acceleration for $v_{\rm Ap0} \gtrsim 0.2$. 
 
The dependence of diagnostic physical quantities on mass loading parameter $\mu$ is presented. Heavily loaded winds correspond to weakly magnetized winds ($v_{\rm Ap0} <1$). The Alfv\'{e}n radius is a decreasing function of $\mu$, while $|B_\phi/B_{\rm p}|_{\rm R_A}$ rises with it. The relations derived in cold Weber \& Davis model are generally obeyed for relatively cold winds in our model, whereas warmer winds show more deviations. The terminal velocity of wind fits perfectly to the derived relation which seems not to be affected by the wind temperature. We deduce that is caused by the nearly adiabatic cooling of the wind. The ratio of Poynting flux to kinetic energy flux toward large radii approaches an asymptotic value of $\sigma\sim2$. The dependance on the field line angular velocity and the adiabatic index are explored as a function of $v_{\rm Ap0}$. The Alfv\'{e}n radius $R_{\rm A}$ and terminal velocity $v_{\rm pf}$ enlarge with larger field line angular velocity $\omega$. The adiabatic index does not seem to impact the wind solution much especially when $v_{\rm Ap0}>1$.

\subsection{Discussion} \label{sec:dis}

\subsubsection{Comparison with Hydrodynamic Model} 

In the first paper of this series, hydrodynamic wind model considering the black hole and galaxy potential is employed to study the wind dynamics toward large distances \citep{cyl19}. The wind solution found in that work requires to pass through the sonic point smoothly, which is the only critical point in the hydrodynamic model. We demonstrate that the relations of physical quantities as a function of cylindrical $R$ are $\rho \propto R^{-2}, v_{\rm p} \propto {\rm const.}$, and $v_\phi \propto R^{-1}$. The wind acceleration is attributed to the conversion of enthalpy to kinetic energy. For hot accretion flows, the radial velocity of wind is nearly constant ($\approx0.2v_{\rm K0}$) with the departure from the wind base.

Including magnetism, we note that the hydrodynamic variables share the same proportionalities with $R$ to those in the hydrodynamic model. The wind is accelerated by both thermal pressure and magnetic force for weak magnetization ($v_{\rm Ap0}\sim0.2$), where the toroidal magnetic pressure gradient dominates the magnetic force. For strong magnetization ($v_{\rm Ap0}>1$), thermal pressure is not important and the acceleration is attributed to magneto-centrifugal force near the surface and magnetic pressure gradient beyond the Alfv\'{e}n radius. This leads to the terminal velocity for the magnetized wind reaching $v_{\rm pf}=0.5v_{\rm K0} (\approx 0.016c)$ for the fiducial model where $v_{\rm Ap0}=0.2$, and boosts to $v_{\rm pf}=v_{\rm K0}(\approx 0.03c)$ and $v_{\rm pf}=5v_{\rm K0}(\approx 0.15c)$ for $v_{\rm Ap0}=1$ and $10$, respectively (Figure \ref{fig:cs}).

\subsubsection{Caveats} 

One limitation of this work is that we do not include galaxy potential in the MHD equations. As the wind propagates over the accretion scales, the gravitational potential from the galaxy will play a role against its outward acceleration. This extra potential shall be involved in the Bernoulli integral. Nevertheless, our hydrodynamic results imply that the galaxy potential does not significantly affect the wind solution when adopting reasonable parameters at the wind launching point. With the complexity of solving for MHD equations, the galaxy potential is thereby temporarily excluded in this work for the sake of simplicity. 

Another caveat comes from the collimation of the wind. To confine the momentum flux of the outflow, it can be either compressed by an external, gas pressure-dominated medium or by the hoop stress associated with the magnetic tension of the toroidal magnetic field. However, the kink instability takes place with the presence of a predominating toroidal field. Once the instability sets in, the collimation provided by the hoop stress is mitigated \citep{eichler1993}. Rather than collimated by the toroidal pinching force, the poloidal disk magnetic field is suggested to preserve the collimation \citep{spruit_etal97}. The winds are expected to experience the confinement via the mechanisms mentioned above. The collimation modifies the trajectory of the wind such that the geometry and strength of poloidal magnetic field would be different. However, previous work have proved that the wind properties are not sensitive to field geometry \citep{bai_etal16}. To properly deal with the collimation of the flow, the force balance in $(R,z)$-plane perpendicular to the poloidal magnetic field should be considered (e.g., \citealp{sakurai1985}; Grad-Shafranov equation). In this case, the solutions obtained for a fixed poloidal magnetic field are still valid, but we should interpret the results in terms of the yet to be determined poloidal field. 
\\ \\
We thank the anonymous referee for the valuable comments on MHD theory. This work is supported in part by the National Key Research and Development Program of China (Grant No. 2016YFA0400704), the Natural Science Foundation of China (grants 11573051, 11633006, 11650110427, 11661161012), the Key Research Program of Frontier Sciences of CAS (No. QYZDJSSW-SYS008), and the Astronomical Big Data Joint Research Center co-founded by the National Astronomical Observatories, Chinese Academy of Sciences and the Alibaba Cloud.

\end{document}